\documentclass[iop]{emulateapj}

\shortauthors{Hoard, Howell, \& Stencel}
\shorttitle{$\epsilon$~Aurigae from the Far-UV to the Mid-IR}

\begin{document}

\submitted{Accepted for publication in the Astrophysical Journal}

\title{Taming the Invisible Monster: 
System Parameter Constraints for $\epsilon$~Aurigae 
from the Far-Ultraviolet to the Mid-Infrared}
 
\author{D. W. Hoard\altaffilmark{1}, 
S. B. Howell\altaffilmark{2}, 
R. E. Stencel\altaffilmark{3}}

\altaffiltext{1}{{\em Spitzer} Science Center, California Institute of Technology, Pasadena, CA 91125}
\altaffiltext{2}{National Optical Astronomy Observatory, Tucson, AZ 85719}
\altaffiltext{3}{Department of Physics and Astronomy, University of Denver, Denver, CO 80208}

\begin{abstract}
We have assembled new {\em Spitzer Space Telescope} Infrared Array Camera 
observations of the mysterious binary star $\epsilon$~Aurigae, along with 
archival far-ultraviolet to mid-infrared data, to form an unprecedented 
spectral energy distribution spanning three orders of magnitude in 
wavelength from 0.1 $\mu$m to 100 $\mu$m.  
The observed spectral energy distribution can be reproduced 
using a three component model consisting of a 
2.2$^{+0.9}_{-0.8}$ $M_{\odot}$ F type post-asymptotic giant branch 
star, and a 5.9$\pm$0.8 $M_{\odot}$ B5$\pm$1 type main sequence 
star that is surrounded by a geometrically thick, but partially transparent, 
disk of gas and dust.  At the nominal HIPPARCOS parallax distance 
of 625 pc, the model normalization yields a radius of $135\pm5$ $R_{\odot}$ 
for the F star, consistent with published interferometric 
observations.  The dusty disk is constrained to be viewed at an 
inclination of $i\gtrsim87^{\circ}$, and has effective temperature 
of $550\pm50$ K with an outer radius of 3.8 AU and a thickness of 0.95 AU.  
The dust content of the disk must be largely confined to grains larger 
than $\sim10$ $\mu$m in order to produce the observed gray 
optical--infrared eclipses and the lack of broad dust emission features 
in the archival {\em Spitzer} mid-infrared spectra.  The total mass 
of the disk, even considering a potential gaseous contribution in 
addition to the dust that produces the observed infrared excess, 
is $\ll1$ $M_{\odot}$.
We discuss evolutionary scenarios for this system that could lead 
to the current status of the stellar components and suggests 
possibilities for its future evolution, as well as potential 
observational tests of our model.\\
\end{abstract}


\section{Introduction}
\label{s:intro}

The bright star $\epsilon$~Aurigae (HD 31964) is a single-lined 
spectroscopic binary that is famous for its long orbital period 
(27.1 yr), which is punctuated by an almost 2 yr long eclipse 
caused by an essentially invisible object \citep{1991ApJ...367..278C}.  
The central problem posed by this system is that if the F star 
component, which dominates the light from the system over a wide 
range of wavelength and is the eclipsed object, is a massive 
supergiant (as its spectrum implies), then the invisible companion 
is surprisingly under-luminous for its mass.  Exotic solutions 
for this mass conundrum involving, for example, a black hole 
\citep{1971Natur.229..178C} are not viable because of the train 
of ever more complicated additional requirements that observational 
constraints impose on such a model.  For example, the lack of 
significant X-ray emission from the system precludes a black hole 
{\em unless} there is no accretion from the disk, which is not 
possible {\em unless} there is yet another unseen body (a massive 
planet, perhaps?) that clears out the space around the black hole, 
and so on (see the discussions in \citealt{1986PASP...98..637B}, 
\citealt{1991ApJ...367..278C}, and \citealt{wolk10}).

By examining the optical spectra of $\epsilon$~Aur near the end of 
its 1954--1956 eclipse, \citet{1959ApJ...129..291H} was able to 
deduce the electron density and develop the hypothesis of a Be 
star-like hot object at the center of a large disk of occulting 
material \citep{1961MmSAI..32..351H}.  \citet{1973ApJ...185..229W} 
reported pioneering infrared (IR) observations that revealed the 
presence of an excess consistent with the disk being a cloud of 
partially ionized gas, with a total projected area comparable to 
that of the F star.   An estimate of electron density from the 
IR excess was consistent with the optical--ultraviolet (UV) 
estimates of $10^{11}$ cm$^{-3}$.

As IR detector technology advanced, \citet{1984ApJ...284..799B} 
reported ground-based IR photometry obtained during the 1982--1984 
eclipse that demonstrated that the excess could be characterized 
as a $T=500\pm150$ K source subtending $8\times10^{-16}$ sr.  
\citet{1985ApJ...299L..99B} refined this result with {\em Infrared 
Astronomy Explorer} (IRAS) satellite photometry during the eclipse, 
extending the wavelength coverage well into the thermal IR, and 
yielding a revised temperature of $475\pm50$ K and angular extent 
of $8.6\pm1.0\times10^{-16}$ sr.  \citet{1985Obs...105...90S} 
examined the same {\em IRAS} data and was led to conclude that 
the disk temperature could be better characterized as a 
$750\pm100$ K source with a projected area about six times that 
of the F star photosphere.  Regardless of its exact characteristics, 
the transiting disk in the $\epsilon$~Aur binary offers a valuable 
opportunity to study its longitudinal structure in ways not possible 
with circumstellar disks around single stars.

We report here on the results from new mid-IR observations of 
$\epsilon$~Aur made with the {\em Spitzer Space Telescope} 
\citep{werner04}, which provide a more precise characterization 
of the occulting body, as well as a new look at archival data at 
shorter wavelengths, which better constrain the stellar components.  
The importance of this exercise lies in the fact that the spectral 
energy distribution (SED) now can be much more precisely defined, 
thanks to the availability of new and recalibrated data spanning 
the far-UV to the mid-IR.  These results strongly imply that the 
putative F supergiant star in $\epsilon$~Aur is more likely a 
lower mass, unstable post-AGB object that previously transferred 
matter to a B(e)-like star companion -- as proposed 
by \citet{webbink85} -- resulting in a complex ``dark matter'' 
disk \citep{2008ApJ...685..418H} that causes the eclipses.

\section{Spectral Energy Distribution -- The Data}

\begin{figure*}[tb]
\epsscale{1.1}
\plotone{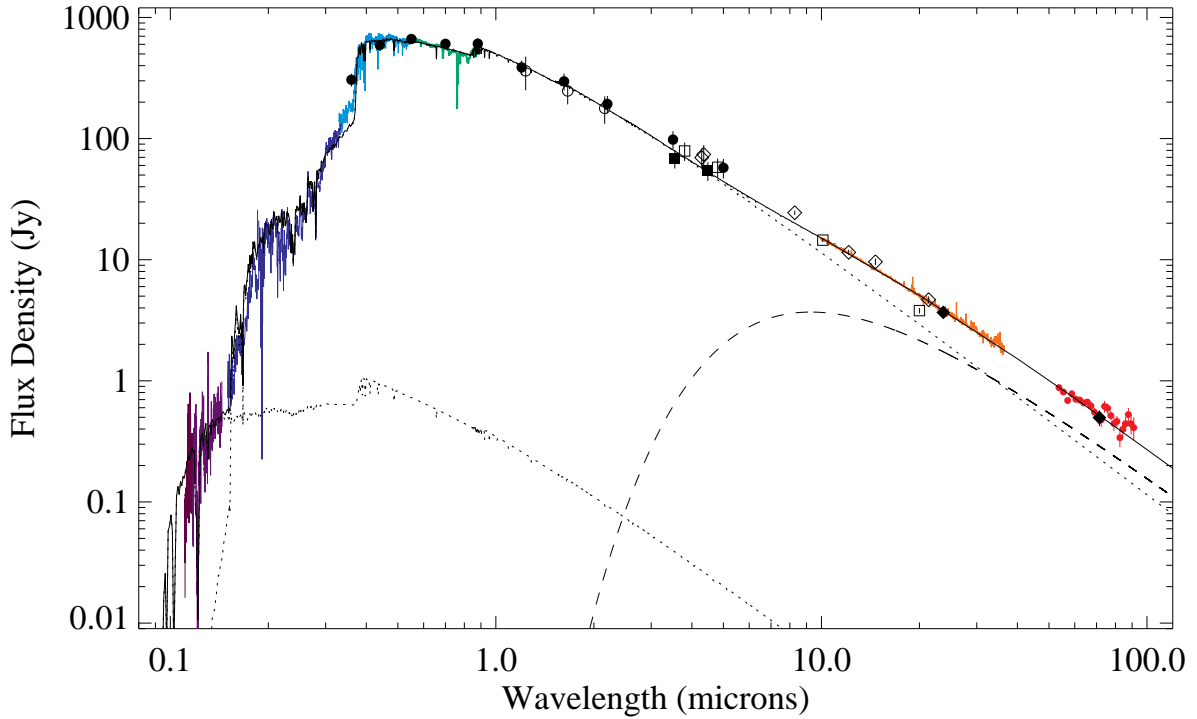}
\epsscale{1.0}
\caption{Observed (dereddened) SED of $\epsilon$~Aur with a three 
component model.  From short to long wavelengths, the photometric 
points are:\ 
U, B, V, R, I from the AAVSO (filled circles), 
J, H, K, L, M (filled circles), 
J, H, Ks from 2MASS (unfilled circles), 
IRAC from {\em Spitzer} (filled squares), 
ground-based L$\prime$, M, N, Q (unfilled squares), 
B1, B2, A, C, D, E bands from {\em MSX} (unfilled diamonds), 
and MIPS from {\em Spitzer} (filled diamonds).  
Vertical error bars are the photometric uncertainties (which 
are dominated by the systematic uncertainty of the dereddening 
process for the dereddened data).  
The spectroscopic data are:\ 
{\em FUSE} (dark purple), 
{\em HST}-GHRS (light purple), 
{\em IUE} (dark blue), 
ground-based optical (light blue and green), 
IRS (orange) 
and MIPS-SED (filled red squares) from {\em Spitzer}.  
See the text and Table \ref{t:data} for more 
information about the data.  The model (solid line) is the sum of 
limb-darkened model F0 (post-AGB) and B5~{\rm V} spectra (dotted lines), 
and a cool blackbody disk (dashed line).  See the text and 
Table \ref{t:model} for more information about the model.
\label{f:sed}}
\end{figure*}

To construct our SED for $\epsilon$~Aur, which spans three orders 
of magnitude in wavelength, we combined data from new observations 
in the mid-IR from {\em Spitzer}, recent observations from the 
American Association of Variable Star Observers (AAVSO) in the 
optical, and archival ground- and space-based observations at other 
wavelengths.  The space-based data in particular come from 
the {\em Midcourse Space Experiment} ({\em MSX}), 
the {\em International Ultraviolet Explorer} ({\em IUE}), 
the {\em Hubble Space Telescope} ({\em HST}), 
and 
the {\em Far Ultraviolet Spectroscopic Explorer} ({\em FUSE}).  
The SED data are listed in 
Table \ref{t:data} and plotted in Figure \ref{f:sed}.  All of the 
observations were obtained outside of eclipse phases; the majority 
were obtained prior to the onset of the eclipse that began in 
late-2009, but well after the end of the prior eclipse in 1984, 
except as noted in Table \ref{t:data}.  We describe our new 
{\em Spitzer} mid-IR observations in more detail below.

%%% BEGIN FIGURE %%%%%%%%%%%%%%%%%%%%%%%%%%%%%%%%%
%%% END FIGURE %%%%%%%%%%%%%%%%%%%%%%%%%%%%%%%%%

\subsection{Spitzer Infrared Array Camera}
\label{s:obs-IRAC}

The expected flux density of $\epsilon$~Aur at 3--5 $\mu$m exceeds 
the nominal saturation limits of channels 1 (3.6 $\mu$m) and 
2 (4.5 $\mu$m) of the Infrared Array Camera (IRAC; \citealt{fazio04}) 
on {\em Spitzer} by factors of several times.  
The rationale for observing $\epsilon$~Aur with IRAC was two-fold:\ 
(1) as a test case for a potential monitoring campaign whose cadence 
could not be accomplished from the ground, and (2) to obtain 
measurements in the 3--5 $\mu$m range whose flux calibration is 
homogenous with that of the other {\em Spitzer} data already obtained 
at longer wavelengths (see Section \ref{s:obs-IRS} and \ref{s:obs-MIPS}).  
As it turned out, a third reason became apparent after the fact; namely, 
that existing older IR data in this wavelength range are not directly 
comparable to the more recent longer wavelength {\em Spitzer} 
data (see discussion in Section \ref{s:disk}).

We utilized an allocation 
of six minutes\footnote{Three minutes of which was the standard 
``slew tax'' applied to all {\em Spitzer} observations.} of 
Director's Discretionary Time in late-April 2009, 
shortly before depletion of the {\em Spitzer} cryogen in mid-May 2009. 
We used a novel observing strategy that took advantage of both the 
extremely short exposure time available in the IRAC sub-array mode 
and the specific placement of the target centroid at the 
intersection of four pixels.  The latter condition spreads 
the total illumination of the detector, 
as well as the brightest part of the point response 
function, over the maximum number of pixels.  
In sub-array mode, observations are obtained in units of 64 exposures 
of a $32\times32$ pixel section of the full $256\times256$ pixel 
array, and are tiled onto the full array for downlink. 
This resulted in four dithered observations of $\epsilon$~Aur 
in each channel, with each dither comprised of 64 consecutive 
0.02-sec exposures. 

We initially applied the IRAC array-location-dependence correction 
to the individual sub-array images, as described in the IRAC Data 
Handbook\footnote{See \url{http://ssc.spitzer.caltech.edu/irac/dh/}.}, 
and then performed aperture photometry on them (e.g., as described 
in \citealt{2009ApJ...693..236H}) using the IRAF\footnote{The Image 
Reduction and Analysis Facility is maintained and distributed by the 
National Optical Astronomy Observatory.} task {\sc phot}.  
We utilized a 10-pixel radius aperture (1 pixel $\approx$ 1.22\arcsec), 
with a 10--20 pixel background annulus, which is the configurations 
for which the corresponding aperture corrections are 1.00 in all 
IRAC channels.

The high time resolution and photon-abundant nature of these 
observations revealed two interesting instrumental artifacts that 
are normally below the detection threshold in {\em Spitzer} IRAC 
observations:\ a rapid ``jitter'' with amplitude of $\approx1$\% 
in the measured flux densities in each sequence of sixty-four 
0.02-sec exposures, as well as overall trends with slopes of 1--2\% 
in measured flux density during each 64-frame sequence.  
The former effect is caused by sub-pixel mismatch between the 
measured and actual centroid of the point response function (PRF), 
propagated through the normal IRAC pixel-phase correction, which 
accounts for the exact position of the target centroid {\em within} 
a pixel.  It is likely caused by resolving the small discrete 
steps in the {\em Spitzer} tracking motion. 

The latter effect appears to be linked to the telescope settle time 
of 2 sec after a slew or dither offset, which is longer than the 
$\approx1.5$-sec total length of one of the 64-frame sub-array 
sequences.  Both of these effects are presumably present 
during the first 2 seconds of any normal, full-array IRAC exposure, 
for which the shortest available exposure time is 2 sec.  
Consequently, in a full-array observation of a fainter target, 
these effects are not time-resolved 
and are below the level of photon counting noise, so go unnoticed 
in the final, single photometric measurement per observation.  
In addition, we see the offsets of order several per cent between 
mean flux densities at different dither positions (caused by 
pixel-to-pixel variations) that are also found in dithered full-array 
observations, and are mitigated by averaging together the flux 
densities measured at the different dither positions.

Additionally, in the channel 2 observations, we see a steep linear 
trend in the measured flux densities in each of the four dithered 
sequences of 64 exposures.  This results in an increase of $\approx$5--7\%, 
with approximately equal slope
from the start to the end of each dither sequence. 
This could possibly include a contribution from the 
so-called ``ramp'' effect noticed in some exoplanet transit light 
curves obtained with IRAC \citep{2009IAUS..253..197D}, but, 
considering the very short total duration of these exposure sequences, 
is more likely due to build-up of latents on the pixels exposed to 
$\epsilon$~Aur during each sub-array sequence.  This particular 
artifact is {\em not} present in our channel 1 data.  We note that 
a 2.5-hr sequence of deep IRAC exposures in the Oph star forming 
region ended about an hour before our $\epsilon$~Aur observation, 
and immediately prior to our observations there was a short (20-min) 
series of observations in the Galactic plane.  So, it is entirely 
possible that the charge traps in channel 1 that would lead to 
latents were already full when our observation started.

The final photometric measurements reported in Table \ref{t:data} 
were obtained as follows:\ for channel 1 (IRAC-1), in order to 
mitigate against one saturated pixel in the $\epsilon$~Aur PRF 
during the first dither sequence, we combined all 256 sub-array 
images using the {\em Spitzer} data analysis software tool MOPEX 
(MOsaicker and Point source EXtractor)\footnote{See 
\url{http://ssc.spitzer.caltech.edu/postbcd/mopex.html}.}, and 
performed aperture photometry (as described above) on the mosaic.  
This results in a flux density that is 0.3 Jy fainter than the 
value obtained by averaging together the aperture photometry results 
for all 256 individual sub-array exposures, including the one 
saturated pixel in 64 exposures.  Qualitatively, this result is 
consistent with the removal of the saturated pixel in 64 of the 
images, resulting in a slightly smaller ``true'' flux density 
measurement.  For channel 2 (IRAC-2), there are no saturated pixels 
in any of the individual sub-array images; however, in order to 
mitigate against the latent charge effect described above, we 
averaged together only the first 5 exposures in each dither sequence.  
In addition, we had to utilize a 5-pixel aperture (with 5-10 pixel 
background annulus) because the 10-pixel aperture was partially 
off the sub-array during two of the dither sequences.  This 
required use of the corresponding aperture correction for 
IRAC-2 of 1.064.

In all cases, as described in the IRAC Data Handbook, we applied 
the pixel-phase correction \citep{hora08} to the photometry in 
both channels.  However, we did not perform the color correction 
other than to utilize the isophotal effective channel wavelengths 
(see Table \ref{t:data}) during our subsequent interpretation of 
the data, which accounts for almost all of the color correction.  
The remaining effect of the color correction is folded into our 
IRAC systematic uncertainty budget.  As described in more detail 
in \citet{2009ApJ...693..236H}, the total uncertainty budget for 
IRAC photometry includes several systematic terms (3\% for the 
absolute gain calibration, 1\% for repeatability, 3\% for the 
absolute flux calibration of the IRAC calibration stars, and 1\% 
for the color correction), as well as a scatter term evaluated 
from the r.m.s.\ scatter of the individual flux density 
measurements after removing all long timescale trends.  
The uncertainties in the IRAC photometry values listed in 
Table \ref{t:data} reflect these terms added in quadrature.

\subsection{Spitzer Infrared Spectrograph}
\label{s:obs-IRS}

A {\em Spitzer} GO-2 observing program (20058; \citealt{stencel07}) 
on $\epsilon$~Aur obtained observations with the Infrared 
Spectrograph (IRS; \citealt{houck04}) and the Multiband Imaging 
Photometer for {\em Spitzer} (MIPS; see below).  For this work, 
we have utilized the current mature pipeline-(re)processed data 
for those observations, and re-extracted final data products.  
The two high resolution~~IRS modules were used during two visits 
to $\epsilon$~Aur,
\clearpage
%%% BEGIN TABLE %%%%%%%%%%%%%%%%%%%%%%%%%%%%%%%%%
\begin{turnpage} % emulate 
\begin{deluxetable}{llllllllll}
\tablewidth{0pt}
\tabletypesize{\scriptsize}
%\rotate %preprint
\tablenum{1}
\tablecaption{The Data\label{t:data}} 
\tablehead{
\colhead{Wavelength} & 
\colhead{Band} & 
\colhead{Source} & 
\multicolumn{2}{c}{Date of Observation:} &
\colhead{Orbital} &
\multicolumn{2}{c}{Flux Density:} &
\colhead{Reference\tablenotemark{d}} &
\colhead{Notes} \\
\colhead{ } & 
\colhead{ } & 
\colhead{ } & 
\colhead{UT} & 
\colhead{JD} & 
\colhead{Phase\tablenotemark{a}} &
\colhead{Observed\tablenotemark{c}} &
\colhead{Dereddened\tablenotemark{d}} &
\colhead{ } & 
\colhead{ } \\
\colhead{($\mu$m)} & 
\colhead{ } & 
\colhead{ } & 
\colhead{(YYMMDD)} & 
\colhead{(JD-2400000)} & 
\colhead{ } & 
\colhead{(Jy)} &
\colhead{(Jy)} &
\colhead{ } & 
\colhead{ } 
}
\startdata
0.111--0.117 & spectrum     & FUSE     & 010107           & 51917          & 0.646        & 0.00025--0.0015     & 0.15--0.3       & \citet{2006ASPC..348..156A} & \nodata \\
0.117--0.146 & spectrum     & HST-GHRS & 960216           & 50130          & 0.466        & 0.002--0.02         & 0.4--1.2        & \citet{1999PASP..111..829S} & \nodata \\
0.150--0.198 & spectrum     & IUE-SWP  & 850203, 850317   & 46099, 46141   & 0.058--0.062 & 0.005--0.8          & 0.6--20         & \citet{1999PASP..111..829S} & (1,2) \\
0.185--0.335 & spectrum     & IUE-LWP  & 861119--861123   & 46753--46757   & 0.124--0.125 & 0.5--48             & 10--130         & \citet{1999PASP..111..829S} & (2,3) \\
0.329--0.549 & spectrum     & ground   & 820405           & 45065          & 0.953        & 25--215             & 150--630        & \citet{1987ApJ...321..450T} & (4) \\
0.567--0.889 & spectrum     & ground   & c.1990--1992     & c.47892--48987 & 0.25--0.33   & 240--330            & 660--520        & \citet{1993PASP..105..693T} & (5) \\
0.360        & U            & AAVSO    & 031101--090701   & 52918--55013   & 0.748--0.959 & 59.1$\pm$1.8        & 306.0$\pm$37.6  & this work                   & (6) \\ % U = 3.870$\pm$0.001
0.44         & B            & AAVSO    & 031101--090701   & 52918--55013   & 0.748--0.959 & 148.4$\pm$4.5       & 591.5$\pm$49.4  & this work                   & (6) \\ % B = 3.611$\pm$0.001
0.55         & V            & AAVSO    & 031101--090701   & 52918--55013   & 0.748--0.959 & 230.4$\pm$6.9       & 663.9$\pm$59.3  & this work                   & (6) \\ % V = 3.038$\pm$0.001
0.70         & R            & AAVSO    & 031101--090701   & 52918--55013   & 0.748--0.959 & 294.6$\pm$8.8       & 605.8$\pm$55.9  & this work                   & (6) \\ % R = 2.498$\pm$0.001
0.88         & I            & AAVSO    & 031101--090701   & 52918--55013   & 0.748--0.959 & 378.4$\pm$11.4      & 606.8$\pm$67.9  & this work                   & (6) \\ % I = 2.107$\pm$0.001
1.20         & J            & ground   & 970907--000418   & 50699--51652   & 0.523--0.620 & 297.1$\pm$10.3      & 386.9$\pm$53.4  & \citet{2001AstL...27..338T} & (7) \\ % J = 1.830$\pm$0.034
1.235        & J            & 2MASS    & 991108           & 51491          & 0.603        & 282.2$\pm$78.6      & 362.5$\pm$112.3 & \citet{2006AJ....131.1163S} & \nodata \\ % J = 1.880$\pm$0.298
1.62         & H            & ground   & 970907--000418   & 50699--51652   & 0.523--0.620 & 254.8$\pm$14.0      & 296.8$\pm$47.5  & \citet{2001AstL...27..338T} & (7) \\ % J = 1.830$\pm$0.034
1.662        & H            & 2MASS    & 991108           & 51491          & 0.603        & 213.6$\pm$35.5      & 247.1$\pm$55.5  & \citet{2006AJ....131.1163S} & \nodata \\ % H = 1.702$\pm$0.178
2.159        & Ks           & 2MASS    & 991108           & 51491          & 0.603        & 162.5$\pm$32.4      & 177.7$\pm$45.4  & \citet{2006AJ....131.1163S} & \nodata \\ % Ks = 1.533$\pm$0.214
2.20         & K            & ground   & 970907--000418   & 50699--51652   & 0.523--0.620 & 177.1$\pm$6.1       & 193.1$\pm$31.7  & \citet{2001AstL...27..338T} & (7) \\ % J = 1.830$\pm$0.034
3.50         & L            & ground   & 970907--000418   & 50699--51652   & 0.523--0.620 & 94.5$\pm$3.3        & 98.0$\pm$16.9   & \citet{2001AstL...27..338T} & (7) \\ % J = 1.830$\pm$0.034
3.544        & IRAC-1       & Spitzer  & 090426           & 54948          & 0.953        & 66.2$\pm$3.0        & 68.6$\pm$12.0   & this work                   & (8) \\
3.8          & L$^{\prime}$ & ground   & 811113           & 44922          & 0.939        & 76.5$\pm$2.6        & 79.0$\pm$13.7   & \citet{1984ApJ...284..799B} & (4,9) \\ % L' = 1.25$\pm$0.02
4.29         & MSX-B1       & MSX      & c.960424--970222 & c.50198--50502 & 0.47--0.50   & 67.9$\pm$6.1        & 69.7$\pm$13.4   & \citet{2003yCat.5114....0E} & \nodata \\
4.35         & MSX-B2       & MSX      & c.960424--970222 & c.50198--50502 & 0.47--0.50   & 72.1$\pm$6.6        & 74.0$\pm$14.3   & \citet{2003yCat.5114....0E} & \nodata \\
4.479        & IRAC-2       & Spitzer  & 090426           & 54948          & 0.953        & 52.9$\pm$2.4        & 54.2$\pm$9.6    & this work                   & (7) \\
4.8          & M            & ground   & 800130--811210   & 44269--44949   & 0.873--0.942 & 56.8$\pm$2.9        & 58.0$\pm$10.4   & \citet{1984ApJ...284..799B} & (4,9) \\ % M = 1.19$\pm$0.02
5.00         & M            & ground   & 970907--000418   & 50699--51652   & 0.523--0.620 & 56.3$\pm$3.4        & 57.4$\pm$10.5   & \citet{2001AstL...27..338T} & (7) \\ % J = 1.830$\pm$0.034
8.28         & MSX-A        & MSX      & c.960424--970222 & c.50198--50502 & 0.47--0.50   & 24.4$\pm$1.0        & \nodata         & \citet{2003yCat.5114....0E} & \nodata \\
9.89--37.14  & IRS-1, -3    & Spitzer  & 051019, 060317   & 53663, 53812   & 0.823, 0.838 & 15.3--5.1           & \nodata         & this work                   & (1,10) \\
10.1         & N            & ground   & 800201--811217   & 44271--44956   & 0.873--0.942 & 14.5$\pm$0.5        & \nodata         & \citet{1984ApJ...284..799B} & (4,9) \\ % N = 1.02$\pm$0.02
12.13        & MSX-C        & MSX      & c.960424--970222 & c.50198--50502 & 0.47--0.50   & 11.5$\pm$0.6        & \nodata         & \citet{2003yCat.5114....0E} & \nodata \\
14.65        & MSX-D        & MSX      & c.960424--970222 & c.50198--50502 & 0.47--0.50   & 9.6$\pm$0.6         & \nodata         & \citet{2003yCat.5114....0E} & \nodata \\
20.0         & Q            & ground   & 811210--811217   & 44949--44956   & 0.942        & 3.8$\pm$0.2         & \nodata         & \citet{1984ApJ...284..799B} & (4,9) \\ % Q = 1.04$\pm$0.04
21.34        & MSX-E        & MSX      & c.960424--970222 & c.50198--50502 & 0.47--0.50   & 4.7$\pm$0.3         & \nodata         & \citet{2003yCat.5114....0E} & \nodata \\
23.675       & MIPS-24      & Spitzer  & 050925, 060223   & 53639, 53790   & 0.820, 0.836 & 3.7$\pm$0.2         & \nodata         & this work                   & (1,11) \\
53.7--91.1   & MIPS-SED     & Spitzer  & 050925, 060223   & 53639, 53790   & 0.820, 0.836 & 0.88--0.41          & \nodata         & this work                   & (1,12,13) \\
71.44        & MIPS-70      & Spitzer  & 050925, 060223   & 53639, 53790   & 0.820, 0.836 & 0.5$\pm$0.07        & \nodata         & this work                   & (1,11) 
\enddata
\tablenotetext{a}{Using the orbital ephemeris JD$_{\rm obs}$ = JD2445525 + 9890E (e.g., \citealt{1985IBVS.2748....1S, 1991ApJ...367..278C}).}
\tablenotetext{b}{See text for a discussion of the dereddening applied to the data.}
\tablenotetext{c}{When necessary, photometric data reported in magnitudes were converted to flux densities using appropriate filter-specific zero point values (e.g., see \citealt{2003AJ....126.1090C, 1988PASP..100.1134B, 1985AJ.....90..896C, 1979PASP...91..589B, 1976ApJ...208..390B, 1963BOTT....3..137M}).}
\tablenotetext{d}{For this work, we have re-extracted all photometric and spectroscopic measurements from archival data, but list here the first original publication of the relevant data.}
\tablecomments{
               (1) Flux densities are the averages of 2 measurements on the indicated dates.
               (2) {\em IUE} exposures SWP 25156 and 25470, and LWP 09565, 09547, and 09554.
               (3) Flux densities are the averages of 3 measurements in the indicated date range.
               (4) Note that the orbital phase refers to the previous cycle (i.e., before the 1982--1984 eclipse).  
               (5) Exact date of spectrum is not given in the corresponding publication.
               (6) Flux densities are the averages of $N$ data points in the indicated date range:\ $N_{\rm U}=411$, $N_{\rm B}=411$, $N_{\rm V}=411$, $N_{\rm R}=37$, $N_{\rm I}=21$.
               (7) Flux densities are the averages of 34 measurements in the indicated date range.
               (8) {\em Spitzer} AOR key 33903360, processing pipeline version S18.7.0.
               (9) Flux densities are the averages of $N$ pre-eclipse data points in the indicated date range:\ $N_{\rm L^{\prime}}=1$, $N_{\rm M}=4$, $N_{\rm N}=4$, $N_{\rm Q}=2$.
               (10) {\em Spitzer} AOR keys 13848832 and 13849600, processing pipeline version S18.7.0.
               (11) {\em Spitzer} AOR keys 13849088 and 13850112, processing pipeline version S16.1.0.
               (12) {\em Spitzer} AOR keys 13849856 and 13849344, processing pipeline version S16.1.0.
               (13) MIPS-SED flux densities have been normalized to the MIPS-70 photometric point by scaling the former by a factor of 0.84.
               }
\end{deluxetable}
\end{turnpage} % emulate 
\clearpage
%%% END TABLE %%%%%%%%%%%%%%%%%%%%%%%%%%%%%%%%%
\noindent and we utilized the {\em Spitzer} data analysis 
software tool SPICE ({\em Spitzer} IRS Custom Extraction)\footnote{See 
\url{http://ssc.spitzer.caltech.edu/postbcd/spice.html}.} to extract 
the calibrated spectra from each visit for each of the two nod 
positions from the combined post-BCD data products (which simply 
co-add the three individual cycle exposures at each nod position 
from each visit).  Module 1 spans 9.9--19.5 $\mu$m, while 
module 3 spans 18.8--37.1 $\mu$m, at a resolving power of $\sim600$ 
in both modules.  Prior to extraction, the {\em Spitzer} data 
analysis software tool IRSCLEAN\footnote{See 
\url{http://ssc.spitzer.caltech.edu/postbcd/irsclean.html}.} was 
used to remove both permanent and rogue hot pixels from the 
combined post-BCD images.  Because of the short exposure times 
used for individual frames (6- and 14-sec in modules 1 and 3, 
respectively), as well as the brightness of the target, no offset 
sky background spectrum was subtracted; this is in keeping with 
the ``best practices'' recommendation for observations of this 
type from the {\em Spitzer} Observer's Manual\footnote{See 
\url{http://ssc.spitzer.caltech.edu/documents/SOM/}.}.  
The four extracted spectra from each module (2 nods from each 
of 2 visits) were then combined in a weighted average after 
rejecting all wavelength points flagged as unreliable by SPICE.

\subsection{Multiband Imaging Photometer for Spitzer}
\label{s:obs-MIPS}

The MIPS \citep{rieke04} data on $\epsilon$~Aur were obtained 
in two observing modes.  The first of these was the normal small 
field photometry mode at 24 and 70 $\mu$m (MIPS-24 and MIPS-70, 
respectively).  We performed aperture photometry on the filtered 
combined post-BCD images following the procedures in the MIPS 
Data Handbook\footnote{See 
\url{http://ssc.spitzer.caltech.edu/mips/dh/index.html}.}; 
operationally, this process is very similar to that used to 
obtain the IRAC photometry.  For the MIPS-24 images, we utilized 
a 35-arcsec aperture with a 40--50-arcsec background annulus, 
and applied the corresponding aperture correction of 1.08.  
For the MIPS-70 images, we utilized a 35-arcsec aperture with 
a 39--65-arcsec background annulus, and applied the corresponding 
aperture correction of 1.22.  As with the IRAC photometry, we 
did not apply the color correction because most of it is accounted 
for by utilizing the isophotal band wavelengths (see 
Table \ref{t:data}).  Instead, we folded the remaining small 
effect into our uncertainty budget.

For MIPS-24 photometry, the total systematic uncertainty budget 
is the quadrature sum of 4\% for the absolute calibration, 
0.4\% for repeatability, 3\% for the color correction, and 
5\% for the aperture correction.  These values reflect upper 
limits from version 3.3 of the MIPS Data Handbook.  In addition, 
the total uncertainty includes a scatter term of 0.006 Jy (0.2\%) 
obtained for the target aperture in the uncertainty (``munc'') 
image provided as part of the post-BCD data products.  
For the MIPS-70 photometry, the corresponding terms in the 
uncertainty budget are:\ 
7\% for the absolute calibration, 
5\% for repeatability, 
3\% for the color correction, 
5\% for the aperture correction, and
a scatter term of 0.007 Jy (1.4\%).

The other MIPS observation utilized the MIPS-SED mode, which 
obtains a very low resolution ($R\approx$15--25) spectrum 
between 52--97 $\mu$m.  We performed a weighted average of 
the post-BCD extracted spectrum from each of the two MIPS-SED 
visits to $\epsilon$~Aur.  The flux calibration uncertainty 
for MIPS-SED mode is $\sim20$\%, and we found it necessary to 
scale the data by a factor of 0.84 (i.e., a 16\% reduction 
compared to the extracted values) in order to match the MIPS-SED 
spectrum with the better calibrated MIPS-70 photometric point.

\section{Spectral Energy Distribution -- The Model}

It has been fairly well established (e.g., see 
\citealt{1965ApJ...141..976H}, \citealt{1971ApandSS..10..332K} 
and the review in \citealt{webbink85}) that the $\epsilon$~Aur 
system consists of three primary components:\ an F star orbited 
by one or two B stars, with a large dusty disk surrounding the 
latter.  It is the disk that eclipses the F star every 27 years.  
The exact details of these components, however, remain rather 
nebulous; for example, the mass of the F star has been proposed 
to be either very high (10--20 $M_{\odot}$; e.g., 
\citealt{1996ApJ...465..371L}) or relatively low (1--4 $M_{\odot}$; e.g., 
\citealt{1996ApJ...465..371L,1987PASJ...39..135S,1986ApandSS.120....1T}).  
The disk around the B star has been variously described as thick 
or thin, flat or twisted, edge-on or slightly inclined, opaque 
or semi-transparent, and ``solid'' or containing a central void.

Fortunately, in the nearly three decades since its last eclipse, 
some important new information about $\epsilon$~Aur has been 
obtained.  This includes a HIPPARCOS trigonometric parallax distance 
(with a nominal value of 625 pc, which we have used in our model 
calculations below; \citealt{1997AandA...323L..49P}), various 
interferometric measurements of the angular size of the F star 
($\approx2.1$--2.3 mas from 0.45 $\mu$m to K-band; 
\citealt{2008ApJ...689L.137S,2003AJ....126.2502M,2001AJ....122.2707N}), 
and the space-based far-UV and mid-IR observations shown in 
Figure \ref{f:sed}, which reveal details about the system 
components that are {\em not} the F star (which otherwise completely 
dominates the SED of $\epsilon$~Aur in the range 0.2--4 $\mu$m).  
Using these data, we are now able to definitively constrain some 
of these heretofore uncertain system parameters.  Our observed 
and model SEDs are shown in Figure \ref{f:sed} and the model 
parameters are listed in Table \ref{t:model}.  We discuss the 
model parameters, constraints, and components in more detail below.

%%% BEGIN TABLE %%%%%%%%%%%%%%%%%%%%%%%%%%%%%%%%%
%\begin{deluxetable}{llll}  % preprint
\begin{deluxetable*}{llll} % emulate 
\tablewidth{0pt}
\tabletypesize{\footnotesize} % emulate
%\tabletypesize{\scriptsize} % preprint
%\rotate
\tablenum{2}
\tablecaption{The Model\label{t:model}} 
\tablehead{
\colhead{Component} & 
\colhead{Parameter} & 
\colhead{Value} & 
\colhead{Reference} 
}
\startdata
System & Adopted Distance, $d$ (pc)    & 625   & HIPPARCOS \citep{1997AandA...323L..49P} \\
       & Inclination, $i$ ($^{\circ}$) & 89 ($\gtrsim87$) & this work, \citet{1996ApJ...465..371L} \\
       & Orbital Separation, $a$ (AU)  & $18.1^{+1.2}_{-1.3}$ & this work \\
       &               &     &               \\
F Star & Spectral Type & F0~{\rm II}--{\rm III}? (post-AGB) & this work \\
       & Temperature, $T_{\rm F}$ (K) & 7750 & this work, \citet{1978AandA....69...23C} \\
       & $\log g$                     & $\lesssim1.0$ & this work, \citet{1978AandA....69...23C} \\
       & Radius, $R_{\rm F}$ ($R_{\odot}$) & $135\pm5$ & this work \\
       & Angular Diameter, $D_{\alpha}$ (mas) & $2.01\pm0.07$ & this work \\
       & Mass, $M_{\rm F}$ ($M_{\odot}$) & 2.2$^{+0.9}_{-0.8}$ & this work \\
       &               &     &               \\
B Star & Spectral Type & B5$\pm$1~{\rm V} & this work \\
       & Temperature, $T_{\rm B}$ (K) & 15,000 & \citet{cox00} \\
       & $\log g$                     & 4.0   & \citet{cox00} \\
       & Radius, $R_{\rm B}$ ($R_{\odot}$) & 3.9$\pm$0.4 & \citet{cox00} \\
       & Mass, $M_{\rm B}$ ($M_{\odot}$) & 5.9$\pm$0.8 & \citet{cox00} \\
       &               &     &               \\
Disk & Temperature, $T_{\rm disk}$ (K) & $550\pm50$ & this work \\
     & Radius, $R_{\rm disk}$ (AU) & $3.8^{+0.1}_{-0.4}$ & this work, \citet{1996ApJ...465..371L} \\
     & Height, $H_{\rm disk}$ (AU) & 0.475 & this work \\
     & Assumed Mass, $M_{\rm disk}$ ($M_{\odot}$) & $\ll1$ & this work \\
     & Inferred Dust Grain Radius, $r_{\rm grain}$ ($\mu$m) & $\gtrsim10$ & this work, \citet{1996ApJ...465..371L} \\
     & Transmissivity Factor & 0.3  & this work \\
     & Emissivity Factor     & 2.43 & this work 
\enddata
%\end{deluxetable}  % preprint
\end{deluxetable*} % emulate 
%%% END TABLE %%%%%%%%%%%%%%%%%%%%%%%%%%%%%%%%%

\subsection{Dereddening of the Observed SED}
\label{s:dereddening}

The observed flux densities were dereddened using the 
UV--optical--IR extinction law from \citet{fitzpatrick07}.  
For $\lambda>5$ $\mu$m, the reddening correction was negligible, 
so was not applied to the data.  An interstellar reddening to 
$\epsilon$~Aur, $E(B-V)=+0.38$, was used for $\lambda=0.33$--$5$ $\mu$m.  
This value was fine-tuned from a starting value of $E(B-V)=+0.35$ 
in the range $E(B-V)\approx0.3$--0.4 found in the literature (e.g., 
\citealt{
2003AJ....126.2502M,
1993PASP..105..693T,
1985Obs...105...90S,
1985NASCP2384...37A,
1979AandA....75..316H,
1978AandA....69...23C,
1969ApJ...157..135H}), 
in order to best match the optical spectra and photometry in the 
SED to the model F star spectrum.  We note that the apparent 
disagreement between the optical spectrum at the shortest wavelengths, 
near 0.3 $\mu$m, most likely reflects issues in accurately 
calibrating a ground-based UV spectrum, rather than an actual 
disagreement with the model, since the model matches the flux 
density of the {\em IUE} spectrum that ends at 0.3 $\mu$m quite 
well.  At wavelengths shortward of $\lambda=0.33$ $\mu$m, 
additional dereddening was applied -- see Section \ref{s:bstar} for details.

% E(B-V) = 0.31(12) -- 2003AJ....126.2502M (from cited Av values using Rv=3.1; Av obtained from Galactic extinction grids of 1992A&A...258..104A.)
% E(B-V) = 0.40    -- 1993PASP..105..693T
% E(B-V) = 0.35    -- 1985Obs...105...90S
% E(B-V) = 0.35    -- 1985NASCP2384...37A (2200A feature)
% E(B-V) = 0.30(5) -- 1979AandA....75..316H
% E(B-V) = 0.39    -- 1978AandA....69...23C
% E(B-V) = 0.34    -- 1969ApJ...157..135H

% E(B-V) = 0.37 -- 2001ApJS..133..345W 

%%% E(B-V) = 0.34    -- 1986PASP...98..637B (repeats 1969ApJ...157..135H)

\subsection{The F Star}

We utilized the $T=7750$ K, $\log g=1.0$ \citep{1978AandA....69...23C}, 
solar abundance model spectrum from the grid of 
\citet{2004astro.ph..5087C}\footnote{The grid of model spectra is 
available at \url{http://wwwuser.oat.ts.astro.it/castelli/grids.html} 
and \url{ftp://ftp.stsci.edu/cdbs/grid/ck04models/}.} to represent 
the F star component in $\epsilon$~Aur, which has the {\em appearance} 
(more on that later) of an F0 supergiant star.

We applied limb-darkening to the model spectrum using the 
\citet{vanhamme93} relation, which is linear in form with 
wavelength-, temperature-, and gravity-dependent coefficients.  
The optical--near-IR region of the SED is dominated by this 
component, and we scaled the limb-darkened model spectrum to 
match the J-band photometric point from \citet{2001AstL...27..338T} 
(see Table \ref{t:data}).  This requires a radius 
of $R_{\rm F} = 135$ $R_{\odot}$, resulting in an angular diameter 
of 2.01 mas.  The radius can only be 
changed by $\pm5$ $R_{\odot}$, corresponding to $\pm0.07$ mas in 
angular diameter, without significantly worsening the match to 
the J point.  Our model angular diameter is somewhat smaller 
than the value of $2.27\pm0.11$ mas measured by 
\citet{2008ApJ...689L.137S} in the K band, but is in agreement 
within $2\sigma$.

The expected surface gravity of a ``normal'' F0 supergiant with 
radius of 135 $R_{\odot}$ and mass of 15--20 $M_{\odot}$ is 
$\log g\approx1.5$, which is one-half dex larger than the value 
selected for our model spectrum for this component.  However, 
as we will demonstrate below, the mass of the F star in 
$\epsilon$~Aur is significantly lower than that of a normal 
F supergiant.  In fact, a more appropriate value for the gravity 
would be $\log g=0.5$; unfortunately, model spectrum calculations 
are apparently not typically made for stars that look like 
F supergiants but have masses an order of magnitude smaller, 
so $\log g=1.0$ is the lowest gravity model available.  
In any case, the gross differences in the model spectra between 
$\log g$ values of 1.0, 1.5, and 2.0 can be compensated for by 
changing the stellar radius by only a few per cent, comparable 
to the determined $\pm5$ $R_{\odot}$ uncertainty, so we 
expect no significant difference in the overall appearance of 
the model SED.

\subsection{The B Star}
\label{s:bstar}

\begin{figure*}[tb]
\epsscale{1.1}
%\plotone{figure2.eps}
\plotone{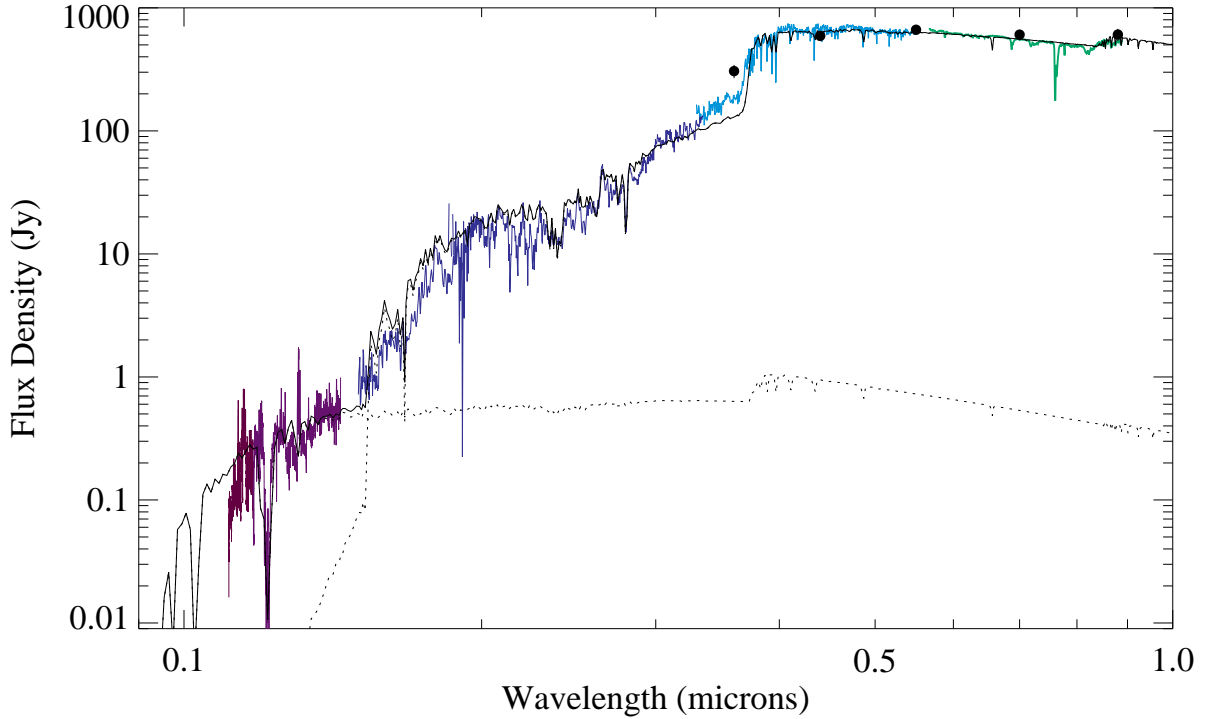}
\epsscale{1.0}
\caption{Expanded view of the ultraviolet wavelength region 
of Figure \ref{f:sed}.
\label{f:uvsed}}
\end{figure*}

Shortward of $\approx0.15$ $\mu$m, the F star spectrum drops off sharply, 
and the B star spectrum dominates the SED of $\epsilon$~Aur.  
This far-UV wavelength region (see Figure \ref{f:uvsed}) 
provides the primary constraint on the spectral type (and, hence, 
mass and radius) of the B star.  Using the interstellar dereddening 
of the observed SED (see Section \ref{s:dereddening}), a B8~{\rm V} 
star (represented by the limb-darkened, 
$T=12,000$ K, $\log g=4.0$, solar abundance model spectrum from 
\citealt{2004astro.ph..5087C}) initially provides a good match 
to the observed SED in this region, with the exception that the 
Ly-$\alpha$ absorption line in the model spectrum is significantly 
broader and deeper than observed in the {\em HST} Goddard High Resolution 
Spectrograph (GHRS) spectrum of $\epsilon$~Aur.

%%% BEGIN FIGURE %%%%%%%%%%%%%%%%%%%%%%%%%%%%%%%%%
%%% END FIGURE %%%%%%%%%%%%%%%%%%%%%%%%%%%%%%%%%

However, the B star is completely embedded in the dusty disk, 
which is viewed close to edge-on and has a significantly larger 
thickness than the radius of any B type main sequence star (see 
Section \ref{s:disk}).  Consequently, we must consider that the 
observed spectral contribution of the B star will necessarily 
have been modified by passage through the disk.  There are two 
likely effects.  The first is that the observed spectral 
contribution of the B star will be overall fainter compared to 
the unobscured case; this is essentially the same thing that 
happens to the F star during eclipse, but affects the B star 
at all times.  In the example mentioned above, the B8~{\rm V} 
star is only a good match to the observed SED {\em if the disk 
is completely transparent}, which, among other inconsistencies, 
would preclude the possibility of the disk eclipsing the F star.

\begin{figure*}[tb]
\epsscale{1.1}
%\plotone{figure3.eps}
\plotone{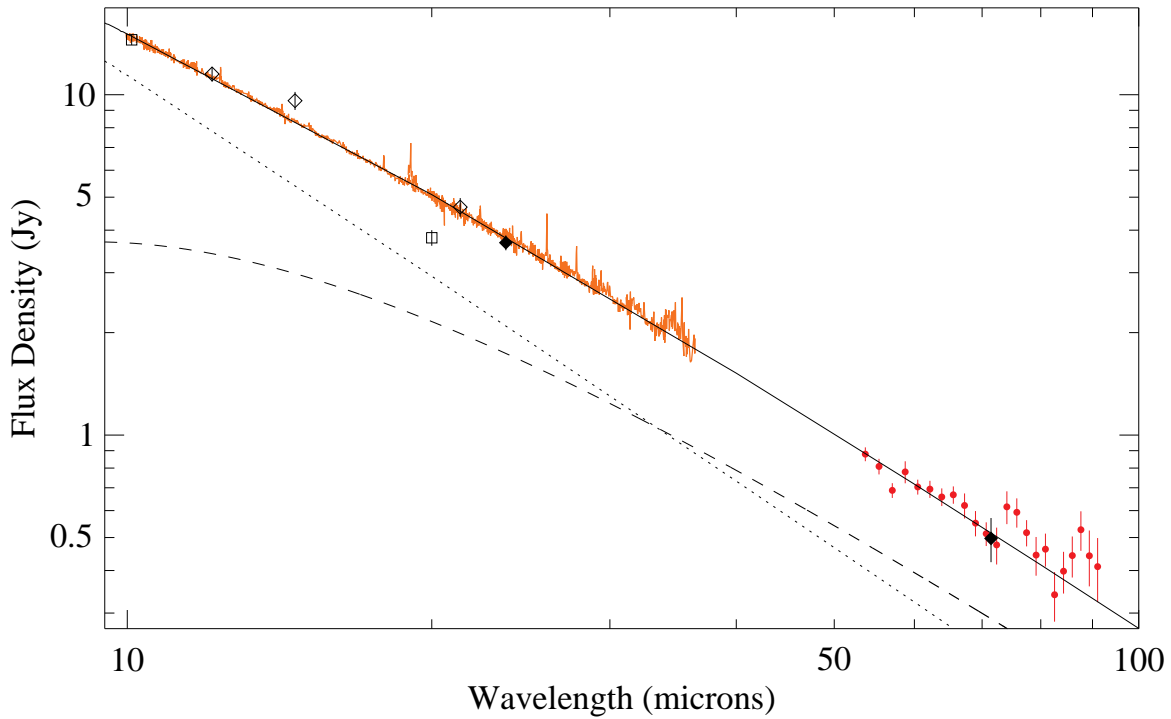}
\epsscale{1.0}
\caption{Expanded view of the mid-infrared wavelength region 
of Figure \ref{f:sed}.
\label{f:irsed}}
\end{figure*}

At optical--near-IR wavelengths, the eclipse of $\epsilon$~Aur 
has been observed to be approximately gray (e.g., 
\citealt{1971ApandSS..10..332K,1996ApJ...465..371L}).  This is 
taken as evidence that the dust grains in the disk around the 
B star must be relatively large, a conclusion that is also 
supported by the lack of broad dust emission features in the 
{\em Spitzer} IRS spectrum of $\epsilon$~Aur (see Figure \ref{f:irsed}), 
which implies 
dust grains of size $\gtrsim10$ $\mu$m (see \citealt{2009ApJ...699.1067M, 
2007prpl.conf..767N, 2006ApJ...638..314D}, and references therein, 
for discussions of the effect of dust grain size on observed 
spectral features).  We can simulate this 
effect by applying a simple scaling factor (which we will refer 
to as the disk ``transmissivity'' factor) to the B star model 
spectrum.  A transmissivity factor of 1.0 would correspond to a 
completely transparent disk, such that the light from an object 
inside or behind the disk would be unaffected.  On the other hand, a 
transmissivity factor of 0.0 would correspond to a completely 
opaque disk, such that the light from an object inside or behind 
the disk would be completely blocked.  For example, if the disk 
attenuates the spectral contribution of the B star by 50\% (a 
transmissivity factor of 0.5), then the observed SED could still 
be matched by the presence of two B8~{\rm V} stars, corresponding to the 
close binary B star model suggested by \citet{1984ApJ...286L..39L} 
and \citet{1985ApJ...288..275E}.

%%% BEGIN FIGURE %%%%%%%%%%%%%%%%%%%%%%%%%%%%%%%%%
%%% END FIGURE %%%%%%%%%%%%%%%%%%%%%%%%%%%%%%%%%

The second likely effect on the observed B star spectrum 
introduced by the circumstellar disk is an additional reddening.
The difficulty of distinguishing between the observable effects 
of interstellar and circumstellar material is often noted 
(e.g., \citealt{2003A&A...408..971H, 1999A&A...350..603H, 
1997ApJ...477..926W}).  In the case of $\epsilon$~Aur, we have 
the relative advantage that the interstellar reddening can be 
determined in the optical region of the spectrum.  In this 
region, the F star dominates the light from the system, and is 
unaffected by the dusty disk (outside of eclipse).  However, 
this still leaves the question of how to account for the 
circumstellar reddening of the B star.

To first order, additional circumstellar reddening of the B star 
might be blamed on the fact that the effect of reddening is an 
order of magnitude greater in the UV than in the optical.  
Consequently, the ``grayness'' of the disk could begin to break 
down at the short wavelengths at which the B star spectrum 
dominates the SED.  There is some disagreement in the literature 
whether this effect is \citep{1986ApandSS.123...31P,1984NASCP2349..365A} 
or is not \citep{1985AandA...144..395F} supported by UV 
observations with {\rm IUE}.

Another plausible scenario is that the hot B star is surrounded 
by a localized shell or torus of gas and small dust grains 
created by the sublimation and photo-spallation of the large dust 
grains that comprise the bulk of the disk (also see 
\citealt{1983ApJ...269L..17C}).  This material would produce an 
additional non-gray attenuation of the B star's spectrum, without 
affecting the bulk properties of the entire disk.  
To account for this, we applied additional dereddening, 
corresponding to a larger value of E(B-V), to just the 
B star-dominated portion of the SED.  We used the 
\citet{fitzpatrick07} reddening law, thereby making the 
implicit assumption that extinction caused by the localized 
material around the B star has the same wavelength dependence 
as that of the interstellar medium (ISM).
We note that this scenario is consistent with the suggested 
presence of a ``hole'' (probably a decrease in optical depth rather 
than an actual absence of disk material) at the center of the 
$\epsilon$~Aur disk, which has been proposed by 
\citet{1971ApJ...170..529W} and others to explain the mid-eclipse 
brightening observed during the 1982--1984 eclipse.  We do not 
further consider the hole since we are not concerned with the 
eclipse behavior, except inasmuch as the typical eclipse depth 
implies that a specific fraction of the F star's light is 
blocked -- see Section \ref{s:diskparams}.

The additional dereddening correction was applied to the 
UV region of the SED as follows.  For $\lambda<0.15$ $\mu$m, 
the SED is completely dominated by the B star component, so we 
simply applied a larger dereddening (i.e., interstellar + circumstellar) 
to these data than to the rest of the SED data (i.e., interstellar only).  
We applied a combined dereddening corresponding to  
$E(B-V)=+0.45$ (reddening steps of +0.05 in $E(B-V)$ between 
$E(B-V)=+0.40$--+0.70 were considered).
For $\lambda=0.15$--$0.33$ $\mu$m, the SED is dominated by neither 
the B star nor the F star component.  In order to construct the 
final dereddened data in this wavelength range (consisting of 
the {\em IUE} spectra), we first created two versions of the data 
dereddened using $E(B-V)=+0.38$ (interstellar only) and $E(B-V)=+0.45$ 
(interstellar + circumstellar), and summed 
them after weighting by the relative contributions of the stellar 
components to the total SED.  For example, at 
$\lambda\approx0.15$ $\mu$m the two stars contribute almost 
equally, whereas at $\lambda\approx0.3$ $\mu$m the B star component 
contributes less than 1\% of the total flux density.  

In the case of two B8~{\rm V} stars (as discussed above), in the 
presence of even a small amount of additional circumstellar reddening, 
corresponding to E(B-V)=0.40 (i.e., an increase of only 0.02 mag 
compared to the interstellar reddening alone), the disk transmissivity 
must be increased to 0.7 to preserve the match to the observed SED.  
For reddening corrections of E(B-V)$\ge$0.45, 
the B8~{\rm V} model continuum slope no longer matches that of the 
observed SED, with the latter being flatter than the former at 
wavelengths $<0.15$ $\mu$m. 
Producing the best fit for these larger circumstellar reddenings, 
which matches the model to the SED at 0.15 $\mu$m, and progressively 
underestimates the observed SED flux density at shorter wavelengths, 
requires a disk transmissivity of $\ge0.95$ for E(B-V)$\ge0.45$, 
respectively.  This is, again, ruled out by the constraint that the 
disk cannot be completely transparent {\em and} eclipse the F star.

However, a single 
B5V star (the next available earlier spectral type, represented 
by the limb-darkened, $T=15,000$ K, 
$\log g=4.0$, solar abundance model from 
\citealt{2004astro.ph..5087C}), combined with a disk transmissivity 
of 0.3, provides a good match to the UV region of the SED.
This includes a much improved match to the width and depth of the 
observed Ly-$\alpha$ absorption line (see Figure \ref{f:uvsed}), 
which is significantly broader and deeper in the B8~{\rm V} model, 
and increasingly too narrow and shallow for B3~{\rm V} and earlier 
spectral type models (see below).

Spectral types of B3~{\rm V} (the next available model template) 
or earlier for the 
B star are not viable solutions (even after considering additional 
attenuation and/or reddening by the disk) because as the B star 
becomes hotter, the continuum slope in this region becomes 
incompatible with the observed (dereddened) SED.  
Specifically, while the observed SED continuum slope rises 
toward longer wavelengths up to 0.15 $\mu$m (at which point the 
B star ceases to dominate the SED), the continuum of a B3~{\rm V} 
star is almost flat and the continuum of a B0~{\rm V} star has 
started to rise slightly toward shorter wavelengths.  
This holds true even after testing by the application of 
implausibly large additional dereddenings.  Hence, the B star 
spectral type is fairly tightly constrained to B5~{\rm V}, 
with an uncertainty of approximately one spectral type in either 
direction.

\subsection{The Dusty Disk}
\label{s:disk}

At a wavelength of $\approx3$--4 $\mu$m, the observed SED of 
$\epsilon$~Aur begins to show 
a deviation from the F star spectrum.  This manifests as an IR 
excess such that, at 100 $\mu$m, the observed SED flux density 
is $\approx3$ times brighter than that of the F star 
alone\footnote{We note that \citet{1994AandA...281..161A} measured 
a 250-GHz (1200-$\mu$m) flux density for $\epsilon$~Aur of $9\pm2$ mJy, 
which is more than an order of magnitude brighter than the 
Rayleigh-Jeans extrapolation of the F star model spectrum at that 
wavelength, $f_{\rm F,1200}=0.8$ mJy.  In fairness, this is also 
a factor of 3 brighter than the sum of the flux densities of the 
F star and blackbody models discussed here at 1200 $\mu$m, which 
is 1.3 mJy.  This possibly implies that there are non-thermal 
emission processes contributing at extremely long wavelengths.}.  

We must draw a distinction between the {\em Spitzer} data, 
which were obtained at orbital phases of $\approx0.8$--0.95, 
and the other mid-IR data (from the ground and {\em MSX}), 
which were obtained at an orbital phase of $\approx0.5$.  
The latter data are systematically somewhat brighter than the 
IRAC data in the $\approx3$--5 $\mu$m region, and show some 
deviations at longer wavelengths, as well.  Some of this behavior, 
particularly the stochastic deviations at the longer wavelengths, 
can likely be attributed to the difficulties in calibrating 
ground-based mid-IR photometry.  
However, the systematic difference at the shorter IR wavelengths 
suggests an actual difference in the characteristics of the cool 
component, either since before the 1982--1984 eclipse (when 
the \citealt{1984ApJ...284..799B} data were obtained), or within 
an orbital cycle since after the last eclipse.
The latter scenario is consistent with viewing the hotter side 
of the disk that is most irradiated by the F star in the data 
from $\phi\approx0.5$, compared to the case in the {\em Spitzer} 
observations, when the opposite, cooler side of the disk is most 
visible.  This effect has been noted in the past 
(e.g., \citealt{2001AstL...27..338T, 1996ApJ...465..371L}), and 
the relevant IR data shown in Figure \ref{f:sed} are consistent 
with a suggested blackbody temperature of 800--1000 K (having 
peak flux density at 5--6 $\mu$m) for the hot side of the disk.  
Future improvements to the disk model would benefit from a 
well-sampled IR data set that spans the entire orbital cycle 
of $\epsilon$~Aur; of course, the long orbital period makes 
this difficult to obtain in practice.  

In the meantime, we have concentrated on reproducing the IR 
excess delineated by the {\em Spitzer} data, since these data 
extend to the longest IR wavelengths in our data set, and have 
a homogenous, well-tested flux calibration.
This IR excess can be reproduced remarkably well with a simple 
single-temperature (550 K) blackbody function, which has peak 
flux density at $\approx9$ $\mu$m.
A $\chi^2$ minimization test applied to 
the data longward of 1 $\mu$m (excluding the data longward of 3 $\mu$m 
that were obtained near orbital phase 0.5 -- see above) 
gives a best range of 500--600 K for the temperature.  
This is the range within which 
the change in $\chi^2$ is less than 10\% compared to the $\chi^2$ 
minimum (which occurs at $T_{\rm bb}=550$ K).  Coincidentally, it 
is also the range 
within which the change in $\chi^2$ in temperature steps of 25 K
is always less than 10\%.  

\citet{1996ApJ...465..371L} calculated a detailed model for the 
disk in $\epsilon$~Aur, but for our purposes, it is not necessary 
to use a model that is so complex.  We acknowledge that our model 
deals with the disk in terms of averaged bulk properties (uniform 
cylindrical volume with a uniform mass distribution) rather than 
more ``realistic'' characteristics such as a specific radial 
density profile, scale height, and so on.  The difficulty of 
specifying the detailed physics of such a disk, in the absence of 
sufficient constraints, is clear from the discussions in 
\citet{1996ApJ...465..371L} and \citet{1991ApJ...367..278C}.  
However, it is also clear that the SED is reproduced very well 
by a parametrically simple model.  Thus, the task of reconciling 
this simple model with detailed dust disk physics awaits even 
more detailed and comprehensive future observational constraints 
(e.g., via interferometric imaging of the disk; \citealt{kloppenborg10}).

In its simplest form, our single-temperature blackbody model for 
the disk is parameterized by a single scaling factor which (in 
conjunction with the distance to $\epsilon$~Aur) is related to the 
projected emitting area of the disk (hence, its size).  Used in 
this fashion, this would correspond to a completely opaque disk 
that emits only from its visible projected surface area.  However, 
as discussed in Section \ref{s:bstar}, we have reason to assume 
that the dusty disk in $\epsilon$~Aur must be partially transmissive. 
As discussed in Section \ref{s:diskparams}, the completely 
opaque case provides only a useful limiting case for the estimation 
of physical parameters of the disk.  
This has led us to express the blackbody component scaling factor 
as the product of two parameters.
The first of these relates to the projected surface area of the disk 
and the second is a disk ``emissivity'' factor.  
The latter parameter scales the blackbody 
flux to account for the emission of dust grains inside the partially 
transmissive disk that are visible from outside the disk and/or 
the fact that, while a true circumstellar disk likely contains a 
range of dust temperatures, we are utilizing only a single-temperature 
model.  
This parameter goes hand in hand with the transmissivity 
factor that we also assigned to the disk (see Section \ref{s:bstar}), 
which specifies how much of the flux of an object behind or inside 
the disk will still be visible through the disk.

Both the transmissivity and emissivity factors are physically 
meaningful, but purely parametric in their application to our 
model.  The former is constrained somewhat by the value required 
to match the far-UV SED with a model B star spectrum.  The value 
of the latter, however, is determined almost solely by minimizing 
the $\chi^2$ of the IR ($\lambda>1$ $\mu$m) region of the 
model.  The caveat to this is that in general the emissivity 
would be 1.0 for a completely opaque disk in which the flux 
contribution from the disk is determined solely by the projected 
surface area -- hence, dimensions and inclination -- of the disk. 
For a more transmissive disk, the value of the emissivity 
factor will be larger than 1.0.

\subsection{Stellar Component Masses}

If the F star in $\epsilon$~Aur is a bona fide F0 supergiant 
(luminosity class {\rm I}), then it must have a mass of 
$M_{\rm F}\approx15$ $M_{\odot}$ or more \citep{cox00}.  
The well known mass function of $\epsilon$~Aur, $f=3.12$ $M_{\odot}$ 
(e.g., \citealt{1996ApJ...465..371L} and references therein), 
then requires that the mass of the B star component be 
13.7 $M_{\odot}$ or more.  This, in turn, would require the 
B star to be of spectral type B1~{\rm V} or earlier, which is excluded 
by the observed far-UV SED, which constrains the 
spectral type to B5$\pm$1~{\rm V} (with a mass 
of 5.9$\pm$0.8 $M_{\odot}$; \citealt{cox00}).  
Among other arguments (e.g., see \citealt{1986PASP...98..389L}), 
a massive evolved B star 
can be excluded as a plausible scenario by considering the 
small relative contribution of the B star to the SED.  
Compared to the factor of $\gtrsim10$ increase in
stellar radius between B5--8 stars of luminosity classes {\rm V} and 
{\rm I} \citep{cox00}, such a star would be quite overluminous 
compared to the observed SED.
Although some additional mass in the B star component can be 
attributed to the dust disk, it seems unlikely that 
$\gtrsim8$ $M_{\odot}$ could be accounted for in this manner 
(see Section \ref{s:diskparams} for confirmation of this).

If the B star is a B5$\pm$1~{\rm V} star, 
and assuming for the moment that the mass of the dust disk is 
negligible in comparison (i.e., $\ll1$ $M_{\odot}$), 
then the known mass function requires that the mass of the 
F star is 2.2$^{+0.9}_{-0.8}$ $M_{\odot}$.  
%%% Adding, for example, an upper limit of one solar mass 
%%% of dust to the B star component raises the allowed mass of the 
%%% F star to 3.3$^{+1.1}_{-0.7}$ $M_{\odot}$.  
This conclusively points to the 
identification of the F star in $\epsilon$~Aur as a low mass 
post-AGB star rather than a normal high mass supergiant 
(\citealt{1985ApJ...288..275E}, \citealt{1987PASJ...39..135S}, 
\citealt{1986ApandSS.120....1T}, and \citealt{1986PASP...98..389L} 
have also suggested low mass F star models for $\epsilon$~Aur, 
although only the last of these has explicitly explored a 
post-AGB identification; also see the review in 
\citealt{2002ASPC..279..121G}).  
Additional support for this conclusion 
is provided in Section \ref{s:evol}.

% M1  M2   a     a1    a2    v1    v2   
% 1.4 5.1  16.84 13.21 3.63  14.53 3.99 -lower limit for B5(1)V
% 2.2 5.9  18.12 13.20 4.92  14.51 5.41 -nominal mass for B5V
% 3.1 6.7  19.31 13.20 6.11  14.52 6.72 -upper limit for B5(1)V
% 2.4 6.1  18.41 13.21 5.20  14.53 5.72 -lower limit for B5(1)V + 1 Msun dust
% 4.4 7.7  20.71 13.18 7.53  14.50 8.28 -upper limit for B5(1)V + 1 Msun dust

\subsection{Disk Size and Mass}
\label{s:diskparams}

There is considerable evidence that both the disk and the orbital 
plane of $\epsilon$~Aur must be viewed close to edge on.  
For example, the flatness of the eclipse profile requires that if 
the eclipsing body is an approximately circular disk, then it must 
be seen in projection at a high inclination (i.e., close to edge on).  
A circular disk viewed in projection at a low inclination (i.e., 
close to face on) would produce a significant variation in the 
geometrically obscured area of the F star as the eclipse progresses.  
In fact, this effect rather strongly constrains the inclination 
to be in the range $i\gtrsim87^{\circ}$ (also see 
\citealt{1996ApJ...465..371L}).

In the limiting case where the inclination is $i=90^{\circ}$, 
the obscuration of the F star during eclipse must be produced 
solely by the disk rim.  We parameterize this with a disk 
thickness, expressed as the height measured from the disk 
mid-plane, $H_{\rm disk}$ (i.e., one-half of the actual geometric 
thickness of the disk).  During eclipse, the optical--near-IR 
flux density is reduced by $\approx$50\%.  To first approximation, 
this would result if the disk completely obscured 50\% of the 
projected area of the F star, which could be accomplished with a 
uniform disk having a thickness of 0.5 AU (i.e., 
$H_{\rm disk}=0.25$ AU).  However, in order to reproduce the 
observed mid-IR SED, under the assumption that the disk is 
completely opaque, would require a disk radius of $\approx20$ AU.

If the masses of the stellar components in $\epsilon$~Aur are 
$M_{\rm F}\approx2.2$ and $M_{\rm B}\approx5.9$, 
then we can calculate the orbital separation of the stellar components 
to be $a\approx18.1$ AU, which clearly rules out a 
disk with a radius of 20 AU.  This orbital separation, in turn, 
allows us to estimate 
the outer radius of the disk (e.g., using equation 2 from 
\citealt{1996ApJ...465..371L}), to be $R_{\rm disk}\approx3.8$ AU.  
Incidentally, the Keplerian orbital velocity at the edge of such 
a disk would be 37 km s$^{-1}$ (starting from the general form of 
Kepler's Third Law, $r^{3} = P^{2}M_{\ast}$ and rearranging to 
yield $v_{\rm rot} = 2\pi \sqrt{M_{\ast}/r}$ for units of distance 
in AU, time in yr, and mass in $M_{\odot}$).  This is comparable 
to the range of $\approx30$--40 km s$^{-1}$ inferred for the 
$\epsilon$~Aur disk via observations of radial velocities of H 
and metal lines (e.g., \citealt{1991AandA...243..230F, 
1987PASJ...39..135S, 1986PASP...98..389L}).

A disk with thickness of 0.95 AU (i.e., $H_{\rm disk} = 0.475$ AU) 
and radius of 3.8 AU, viewed at an inclination of $89^{\circ}$, 
is sufficient to produce a geometric obscuration of the F star's 
disk of 72\% (i.e., 28\% of the F star is completely unobscured 
by the disk during eclipse).  A disk transmissivity of 0.3 (see 
Section \ref{s:bstar}) allows 30\% of the flux from the obscured 
portion of the F star (i.e., 22\% of the total flux) to be visible 
through the disk, resulting in the 50\% value constrained by the 
eclipse depth.  Finally, an emissivity factor of 2.43 is required 
to produce the model shown in Figure \ref{f:sed} (i.e., the observed 
flux of the slightly transparent disk is a factor of 2.43 larger 
than would be observed if only the opaque outer surface of the 
disk contributed).  Additional solutions at other inclinations 
(always $\gtrsim87^{\circ}$) are possible for other combinations 
of disk height, transmissivity, and emissivity (e.g., 
Figure \ref{f:sed} shows the model produced by the disk parameters 
listed in Table \ref{t:model}; however, with $r_{\rm disk}$ fixed 
at 3.8 AU and transmissivity fixed at 0.3, identical eclipsed 
fraction and model SED are achieved using $i=88^{\circ}$, 
$H_{\rm disk}=0.725$ AU, and emissivity of 1.54, or using 
$i=90^{\circ}$, $H_{\rm disk}=0.375$, and emissivity of 3.41).  
As noted in \citet{1973Ap&SS..21..263H,1974ApJ...189..485H}, 
the uniqueness of disk models for $\epsilon$~Aur is problematic.

Up to this point, we have assumed that the mass of the dust 
disk is negligible compared to the stars in $\epsilon$~Aur.  
Let us now test the validity of that assumption, starting 
with comparisons to other astrophysical disks.  For example, 
the dust disks around young (T Tauri) stars are found to have 
a sharply peaked distribution of total (gas + dust) mass 
centered on 0.01 $M_{\odot}$, with the majority of objects 
in the range 0.001--0.1 $M_{\odot}$ \citep{1998apsf.book.....H}.  
However, these disks also have characteristic sizes (outer radii) 
of 50--100 AU or more, implying that the $\epsilon$~Aur disk, 
if similar in structure, would likely be more than two orders 
of magnitude less massive (much less if we consider only the 
mass of dust).  

\citet{2009ApJ...699.1067M} used {\em Spitzer} observations 
of 52 A and late-B type main sequence stars to estimate minimum 
dust disk masses of up to 0.6 $M_{\rm moon}$ 
($\approx2\times10^{-8}$ $M_{\odot}$).  The minimum disk masses 
for most of their sample were several orders of magnitude 
smaller.  They note that the total disk masses (i.e., including 
larger items that evade detection), assuming parent body 
planetesimal sizes of 1 km, could be a factor of $10^{4}$ 
larger (i.e., $\sim10^{-4}$ $M_{\odot}$).  They modelled these 
disks as annular rings around the parent star, and found a 
median radius of 11.9 AU (range of 5--93 AU) with a radial 
spread of $\sim3$--40 AU.  This would present a surface area 
larger by a factor of at least $\sim5$ than the disk in 
$\epsilon$~Aur, so we would expect the disk masses from 
\citet{2009ApJ...699.1067M} to overestimate the $\epsilon$~Aur 
disk mass, if they are otherwise comparable in structure.

Another estimate of the $\epsilon$~Aur disk mass can be made 
by comparison to the circumbinary dust 
disk that produces the mid-IR excess in the cataclysmic variable 
V592~Cassiopeiae.
This disk, which is itself the most massive dust disk yet known for 
a cataclysmic variable, 
has been modeled to contain 
$2.3\times10^{21}$ g ($\approx10^{-12}$ $M_{\odot}$) 
of dust \citep{2009ApJ...693..236H}.  
If we scale up the V592~Cas dust disk to the same volume as the 
$\epsilon$~Aur disk, then the mass of dust in the latter would 
be $\sim6\times10^{-5}$ $M_{\odot}$.  It is not clear if this 
scaling up exercise is appropriate, since the morphology and, 
hence, presumed mass distribution of the circumbinary disk in 
V592~Cas is quite different (geometrically very thin compared to 
its radius, \`{a} la the rings of Saturn) compared to the 
$\epsilon$~Aur disk (which is quite thick relative to its radius, 
almost toroidal).  However, the binary star hosts of these 
disks are similar in all but scale, so we might expect a 
scaled comparison of their dust disks to still provide some 
useful limiting values.

From opacity arguments, \citet{1987ApJ...315..296H} and 
\citet{1984ApJ...284..799B} estimated disk masses of 
$\sim10^{-6}$ $M_{\odot}$ (gas only) and $\sim10^{-7}$ $M_{\odot}$ 
(gas and dust), respectively.  We can attempt to further quantify 
an upper limit to the disk mass in $\epsilon$~Aur in a similar fashion.  
We start by assuming that the 
dust in the disk is comprised solely of spherical 
silicate grains with density of 3.0 g cm$^{-3}$ and radius of 
10 $\mu$m.  If we imagine looking through a 1 cm$^2$ column of 
such grains, then on average 
$[(1 \, {\rm cm}) / (0.001\sqrt{2} \, {\rm cm \, grain}^{-1})]^2 = 5\times10^5$ 
dust grains per cm$^2$ would be required to produce complete 
geometric obscuration along the length of the column (i.e., any 
photon traveling along the column will inevitably be intercepted 
by a dust grain).  The factor of $\sqrt{2}$ provides for the round 
cross-sections of the spherical dust grains to overlap in order 
to completely fill a 1 cm$^2$ square column cross-section.  
If the column length is of the order of the disk radius, then the 
mean dust grain number density is $8.8\times10^{-9}$ cm$^{-3}$, 
corresponding to a total mass of dust in the disk of 
$\approx8\times10^{-9}$ $M_{\odot}$.  Of course there could be 
{\em more} mass than this -- we cannot discriminate past the 
point at which the disk would be completely opaque due to the 
column density of dust grains.  However, the fact that we do 
see the B star inside the disk implies that (as already assumed 
in our model), the disk is not completely opaque, so this mass 
estimate is necessarily an upper limit.  

The mass of a spherical dust grain with a fixed density increases 
in proportion to $r_{\rm dust}^{3}$, while the cross-section of 
the grain increases as $r_{\rm dust}^{2}$.  Thus, increasing the 
grain radius by a factor of 10 decreases both the requisite column 
density to achieve total obscuration and the total number 
of dust grains in the disk by a factor of $10^{2}$, but increases 
the total mass of the grains only by a factor of 10.  In this manner, 
we can estimate that if the total mass of dust in the disk is 
$\sim1$ $M_{\odot}$ (and the disk is completely opaque), then 
the ``dust grains'' would have to have implausibly large 
radii, $r_{\rm dust}\sim1$ km.

Therefore, based on all of the comparisons and calculations 
described above, it appears safe to say that the mass of the 
dust disk in $\epsilon$~Aur is $\ll1$ $M_{\odot}$, with the 
range of estimates topping out at around one-hundredth of a 
per cent of a solar mass.
There is presumably also gas in the disk, but even 
at the ``standard'' ISM gas-to-dust mass ratio of $\sim$100:1 
\citep{tielens05}, the mass contribution from the gas will still 
leave the estimated total disk mass at well under 1 $M_{\odot}$.
Plausibly large dust grains (e.g., centimeter-scale ``pebbles'') 
are expected to have distinct 
sub-mm emissivity properties (perhaps related to the 9 mJy 250-GHz 
flux density found by \citealt{1994AandA...281..161A}), and deserve 
further study at long wavelengths.

\section{Evolutionary Models for Epsilon~Aur}
\label{s:evol}

Just after the last eclipse event of $\epsilon$Aur, \citet{webbink85} 
presented a few model scenarios for the evolution of the binary.  
Given the poorly known distance at the time and its MK spectral 
classification, $\epsilon$~Aur was considered an F supergiant 
based on its spectral characteristics (e.g., narrow lines) and 
effective temperature.  \citet{1978AandA....69...23C} performed 
a spectroscopic fine analysis to derive $\log g=1$, indicative 
of a low surface gravity consistent with a very large star.  
\citet{webbink85} assumed three possible absolute luminosities 
for the supergiant, bracketing the full range of the distance 
estimate at the time.  His models for $\epsilon$~Aur initially 
included a pre-main sequence star, making the binary extremely 
young, but then he quickly argues against such a scenario on a 
number of grounds.  Next, \citet{webbink85} considered a number 
of shell and core burning scenarios, all more or less suggestive 
that $\epsilon$~Aur was a post-main sequence star located in one 
or another ``loop'' of its evolution crossing the H-R diagram.  
Some of these loops involve significant mass loss from the star 
during an earlier stage.

While the general idea of the presence of a post-main sequence 
evolved F star in $\epsilon$~Aur is not a new idea (e.g., see 
\citealt{1986PASP...98..389L}), we have proposed here that the 
apparent F supergiant star is, in fact, a bright post-asymptotic 
giant branch (post-AGB) star that started on the main sequence 
as a $\gtrsim7$ $M_{\odot}$ star.  
The observed $^{12}$C/$^{13}$C ratio of $\sim10$ from CO line 
observations \citep{1987ApJ...315..296H} and possibly elevated 
barium abundance of $2\times$ solar \citep{1978AandA....69...23C} are 
indicative of AGB thermal dredge-up and s-process enhancement 
appropriate for a post-AGB star.  
We note that the elevated barium abundance result from 
\citep{1978AandA....69...23C} would benefit from confirmation 
using updated oscillator strengths, and in comparison to line 
strengths of other s-process elements.
We believe that the mass of the 
F star is currently near 2.2 $M_{\odot}$, while its size is 
that of a supergiant (135 $R_{\odot}$).  The constraints on the 
mass and radius of the F star, which include comparison with 
multi-wavelength (i.e., the SED presented here) and 
interferometric (e.g., \citealt{2008ApJ...689L.137S}) observations 
and the HIPPARCOS parallax distance, as well as known kinematic 
properties of $\epsilon$~Aur, are discussed in detail above.  
We justify the plausibility of the proposed low mass, post-AGB 
F star below.

The study of post-AGB stars is a developing field in stellar 
evolution research, as more examples are identified 
observationally (see \citealt{2007yCat..34690799S}).  
Among the few single star evolutionary models that come close 
to the post-AGB phase, are those presented by 
\citet{2007AandA...469..239M}, which include thermal pulse 
tracks for stars up to 5 $M_{\odot}$ and a range of metallicity.  
The inherent instabilities limit the full exploration of these 
extreme late phases of evolution, but such work provides 
guidance for interpreting the F star in $\epsilon$~Aur.  
Given the observational constraints on the F star, namely 
$T_{\rm eff}$, $R/R_{\odot}$, $\log L = 4.7$, $Z\sim Z_{\odot}$ 
and the $\sim100$~d quasi-period of photospheric variations, 
the $5$ $M_{\odot}$ tracks in \citet{2007AandA...469..239M} 
come closest to approaching these values and have evolutionary 
timescales approaching $10^{5}$ yr.  By extension, we could 
place the progenitor of the F star in the 6--8 $M_{\odot}$ 
range, making it a candidate for so-called super-AGB status 
\citep{2009MNRAS.395.1409S}.  
Tidal interaction between components in $\epsilon$~Aur might 
account for the inflated size of the F star and some of the 
out-of-eclipse light variations.

The effective temperature and luminosity of the F star component 
in $\epsilon$~Aur matches well with an initially 7 $M_{\odot}$, 
post-AGB star early in its evolution away from the AGB, probably 
after being ``born-again'' by a thermal pulse due to rapid core burning.  
However, the evolution tracks in the H-R diagram for 
the initial movement away from the AGB and following a thermal 
pulse are indistinguishable, so we can not tell where in its 
post-AGB evolution the F star is actually located.  In any case, 
these ``loops'' on the H-R diagram are extremely rapid for such 
a star, with durations of only a few thousand years 
\citep{1995AandA...299..755B}.  Ancient star catalogs, however, 
do not imply much change in the appearance of $\epsilon$~Aur on 
the timescale of recorded human history 
\citep{1991ApJ...367..278C,2002ASPC..279..121G}.

The known mass function of $\epsilon$~Aur requires that the F star
currently has a mass of $\sim$1--3 $M_{\odot}$ for a $\sim$5--7 
$M_{\odot}$ B star (where the range in B star mass represents the 
constrained range of its spectral type, B5$\pm$1~{\rm V}).
By implication, during the past few 100,000 years, the F star must 
have lost $\sim5$ or more solar masses of material during its AGB 
and post-AGB evolution.  This high mass loss amount argues for the 
star being in its first or even second thermal pulse loop tour of 
the upper H-R diagram. The F star was the initially more massive 
star in the $\epsilon$~Aur binary system, starting on the main 
sequence as an early B star.  This assignment gives $\epsilon$~Aur 
an age of approximately 60--80 million years since the ZAMS.

The additional complication of binary star evolution requires 
further work (see \citealt{2008MNRAS.384.1109E}), but assessing 
the F star to be in a transitional state (blue loop) in its 
evolution makes an attractive and testable hypothesis.  
The implied $10^{4}$ yr or longer timescales could provide 
ample time to transfer and lose mass in and around the binary, 
but not allow the resultant disk to dissipate, nor the 
photosphere to chemically modify due to hot bottom burning.
The B5~{\rm V} star and its surrounding dust disk are likely capture 
sites for some of the mass lost by the present-day F star, but 
not as much as one might think. As the precursor F star expanded 
after its main sequence life was over, the binary nature of the 
system would have allowed matter to flow through not only the 
inner Lagrange point (L1, located between the two stars) but 
also a greater flow would have escaped through the outer Lagrange 
point (L2), as modeled and discussed in \citet{2002ApJ...575..461E} 
in relation to Algol-like binaries. The issue of large amounts of 
mass loss in close binaries, especially the idea of non-conservative 
mass loss through the outer Lagrange points, is famously known as 
the Algol Paradox and was treated in detail in 
\citet{1978ASSL...68.....K}.  

If the mass lost from $\epsilon$~Aur 
leaked out of the L2 side of the binary, then it might 
remain ``hidden'' in a thin, flat ring of measurable extent.  
Observers might look for evidence of faint, tell-tale circumbinary 
emission from such residual material around $\epsilon$~Aur.  
Taking the advanced post-AGB evolution nature of the primary star, 
we believe that the ``supergiant'' size is due to continual mass 
loss via a slow expanding photosphere heading toward shell ejection 
as in a planetary nebula or similar type of object (e.g., Sakurai's 
object or FG~Sagittae; \citealt{2003ApJ...583..913L}).  The inferred 
F star radius of 135 $R_{\odot}$ is consistent with these stars 
and the models produced for them.

\section{Conclusions}

We have analyzed an unprecedented SED of $\epsilon$~Aur, 
assembled from observational data spanning three orders of magnitude 
in wavelength.  
In conjunction with constraints provided by other published 
information about this enigmatic binary star, such as its orbital dynamics, 
we can conclude that:\
(1) the F star component is a low mass post-AGB object, 
(2) the B star component is a B5$\pm$1~{\rm V} star, and 
(3) the dusty disk is a partially transparent, low mass disk of 
predominantly 10 $\mu$m or larger grains.  
The identification of 
the F star in $\epsilon$~Aur as a normal high mass F supergiant is 
simply no longer tenable as a plausible scenario. The requisite 
mass of the B star plus disk cannot be made to satisfy the well-known 
mass function for this system (without invoking exotic scenarios 
involving compact, non-luminous sources of 5--10 $M_{\odot}$ of 
additional mass that do not produce X-rays -- see Section \ref{s:intro}), 
while at the same time satisfying the constraints on the luminosity 
and SED by the data we have 
assembled here.  From an evolutionary standpoint, the F star must 
have been initially the more massive of the stellar components in 
the binary, and some or all of the mass in the dusty disk around 
the B star may have been transferred from the F star precursor.  
The bulk of the mass is likely to have 
escaped from the binary during 
the evolution of the F star precursor, and might be visible in 
sensitive observations at far-IR wavelengths.  As a post-AGB object, 
the F star is in a relatively rapid transitional phase of stellar 
evolution, and we should expect significant changes in the appearance 
of $\epsilon$~Aur after the next few thousand to tens of thousands 
of years.  In the meantime, observational studies of the out-of-eclipse 
disk at wavelengths greater than 30 $\mu$m and via 
interferometric imaging are to be encouraged.

\acknowledgements{We thank Sally Seebode for her valuable assistance 
in assembling the data used in this work, and the anonymous referee, 
whose comments and suggestions improved the manuscript.  
D.\ W.\ H.\ thanks S.\ Carey 
and J.\ Surace of the IRAC Instrument Support Team for helpful 
discussions about the properties of the IRAC detector hardware.  
We give a special thanks to P.\ Kohler \citep{1992CandE...267...34K} 
for devising an appropriate epithet for $\epsilon$~Aur 
(``le monstre invisible''), which we have borrowed for the title 
of our paper.  We acknowledge with thanks the variable star 
observations from the AAVSO International Database contributed 
by observers worldwide and used in this research.  This work is 
based in part on observations made with the {\em Spitzer Space 
Telescope}, which is operated by the Jet Propulsion Laboratory, 
California Institute of Technology, under a contract with the 
National Aeronautics and Space Administration (NASA).  Support 
for this work was provided by NASA.  Some of this work was also 
supported by NASA through contract agreement 1275955 with the 
University of Denver, issued by the Jet Propulsion Laboratory, 
California Institute of Technology.  R.\ E.\ S.\ is grateful to 
the estate of William Herschel Womble for support of astronomy 
at the University of Denver.  This work makes use of data products 
from the Two Micron All Sky Survey, which is a joint project of 
the University of Massachusetts and the Infrared Processing and 
Analysis Center/Caltech, funded by NASA and the NSF.  Some of the 
data presented in this paper were obtained from the Multimission 
Archive at the Space Telescope Science Institute (MAST). STScI is 
operated by the Association of Universities for Research in 
Astronomy, Inc., under NASA contract NAS5-26555. Support for 
MAST for non-HST data is provided by the NASA Office of Space 
Science via grant NAG5-7584 and by other grants and contracts.  
This research has made use of the SIMBAD database, operated at 
CDS, Strasbourg, France, and NASA's Astrophysics Data System.\\

{\it Facilities:} 
\facility{FUSE}, 
\facility{HST (GHRS)}, 
\facility{IUE}, 
\facility{AAVSO}, 
\facility{IRAS}, 
\facility{Spitzer (IRAC, IRS, MIPS)}}

\end{document}